\documentclass[
reprint,
showpacs,preprintnumbers,
amsmath,amssymb,
aps,
pre,
  superscriptaddress
]{revtex4-1}
\usepackage{soul}
\usepackage{graphicx}
\usepackage{dcolumn}
\usepackage{bm}
\usepackage{braket}
\usepackage{color}
\usepackage{caption}
\usepackage{graphicx}
\usepackage{ragged2e}
\usepackage{natbib}
\usepackage{hyperref}
\usepackage{subcaption}
\usepackage{url}
\usepackage[utf8]{inputenc}
\usepackage{soul}
\usepackage{amsmath}
\usepackage{comment}
\usepackage{color}
\usepackage{subcaption}
\DeclareCaptionFormat{myformat}{#1#2(Color online) #3}
\DeclareCaptionJustification{justified}{\justifying} 
{\captionsetup{format=myformat,labelsep=period,justification=justified, font=small}
\figure
}%
{\endfigure}%

\begin{document}
\preprint{APS/123-QED}
\title{The Avian Compass can be Sensitive even without Sustained Quantum Coherence
} 
\author{Rakshit Jain}
\affiliation{Department of Physics, Indian Institute of Technology Bombay, Powai, Mumbai}
\author{Vishvendra Singh Poonia}
\author{Kasturi Saha}
\author{Dipankar Saha}
\author{Swaroop Ganguly}
\email{sganguly@ee.iitb.ac.in}
\affiliation{Department of Electrical Engineering, Indian Institute of Technology Bombay, Powai, Mumbai }


\date{\today}%

\begin{abstract}
Theoretical studies indicating the presence of long-lived coherence in the radical pair system have engendered questions about the utilitarian role of sustained coherence in the avian compass. 
In this manuscript, we investigate this for a realistic multi-nuclear radical pair system, along with the related question of its sensitivity to the geomagnetic field.
Firstly, we find that sustenance of long-lived coherence is unlikely in a realistic hyperfine environment. 	Secondly, probing the role of the hyperfine interactions on the
compass sensitivity, we establish the hyperfine anisotropy as an essential parameter for the sensitivity. Thereby, we are able to identify a parameter regime where the compass would exhibit sensitivity even without sustained coherence.
\end{abstract}


\maketitle
\section{Introduction}
\label{Intro}
Certain biological systems seem to sustain and utilize quantum effects under ambient conditions~\citep{engel2007evidence,panitchayangkoon2010long,ishizaki2010quantum,ritz2000model,gauger2011sustained}.
They have sparked a lot of interest because understanding their functionality could facilitate the exploitation of quantum effects like coherence and entanglement for technological applications like computing, communication, and sensing. Avian magnetoreception is thought to be one such `quantum biological' phenomenon in which migratory birds navigate long distances by sensing the geomagnetic field~\cite{gauger2011sustained}.

Magnetoreception in migratory birds has been investigated through behavioral experiments ~\citep{wiltschko1972magnetic,wiltschko2005magnetic,wiltschko2002magnetic,ritz2004resonance,wiltschko2006avian,ritz2009magnetic}. These have shown that avian magnetoreception has certain characteristic features, viz. a) photo-initiation by a certain frequency of light~\citep{muheim2002magnetic,wiltschko2005magnetic}, b) dependence on only the inclination of the geomagnetic field ~\citep{wiltschko1972magnetic}, c) disruption by radio-frequency fields of certain frequencies~\citep{thalau2005magnetic,wiltschko2015magnetoreception,ritz2004resonance,ritz2009magnetic}, d) an adaptive selectivity around the local geomagnetic field intensity, also known as the `functional window' property~\citep{wiltschko2002magnetic,wiltschko2006avian}. Two hypotheses were proposed to explain these behavioral characteristics, namely the magnetite particle model and the radical pair (RP) model. 
The magnetite hypothesis suggests the geomagnetic field is sensed by magnetite particles acting as `compass needles', while the RP model proposes a chemical compass, wherein the geomagnetic field influences the spin dynamics of a pair of photo-generated radicals in the bird's retina.
The evidence from the behavioral experiments~\citep{schulten1978biomagnetic,ritz2000model,ritz2004resonance} seems to strongly favor the RP model. This model involves the formation of a pair of radicals by a photon of appropriate frequency, with each radical having an unpaired electron. The spin of the electron pair on these radicals interact with their local hyperfine environment along with the local geomagnetic (Zeeman) field before the radicals recombine back. These interactions cause the initial state of the radical pair to evolve before recombination. Under this evolution, the electron spins might undergo decoherence either due to environmental noise or hyperfine interaction with the nearby nuclear spins \cite{tiersch2012decoherence}.
The chemical product after the recombination depends on the spin state of the radicals just before recombination, thus yielding two distinguishable products after recombination, namely a singlet product and a triplet product resulting from the singlet and triplet radical pair precursors respectively. These ratio of the recombination products contains information about the local geomagnetic field inclination, which is decoded by the avian neural system and used to aid navigation.

The RP model has been considerably successful in explaining most of behavioral characteristics of avian compass~\citep{ritz2000model,gauger2011sustained,bandyopadhyay2012quantum,xu2013estimating,poonia2017functional}. However, there are two important aspects of the spin dynamics of radical pair that have not been examined rigorously till date, namely: a) Are quantum effects like coherence and entanglement sustained in this system under biological conditions? b) If so, do they play an utilitarian role?
Gauger et al. calculated the radical pair lifetime (coherence time) in order for the single-nucleus radical pair mechanism to be sensitive to the geomagnetic field and concluded the spin superposition needs to be sustained for tens of microseconds~\citep{gauger2011sustained}. They also studied the entanglement dynamics in the RP system. Bandyopadhyay et al. strengthened these claims and reported that the candidate molecule responsible for the generation of radical pairs (cryptochrome) too has the lifetime of the same order
~\citep{gauger2011sustained,bandyopadhyay2012quantum}. 
Further, Cai and Plenio analyzed the chemical compass based on its analogy with the quantum interferometer and analyzed the global coherence (coherence of the electron pair + nuclei system) for a large sample set of radical pair systems and concluded that the coherence is a \textit{resource} for the system~\citep{cai2013chemical}. However, the role of electron pair coherence (termed as local coherence by them) is meager as compared to the global coherence~\citep{cai2013chemical}. Their result seems to make the case that statistically global coherence might be enhancing the sensitivity of the compass. Kominis and co-workers too suggested along the similar lines that coherence is indeed a resource for the avian compass~\citep{katsoprinakis2010coherent}. Further work on the radical pair mechanism analyzed the role played by the nuclear hyperfine interaction in the magnetoreception~\citep{steiner1989magnetic,ritz2000model,tiersch2012decoherence,poonia2015state,timmel2001model,timmel2004study}. 
It concluded that the anisotropy in the nuclear hyperfine interaction is indispensable~\citep{timmel2001model,cintolesi2003anisotropic,hogben2012entanglement,tiersch2012decoherence,poonia2015state} for the compass operation.  

Now, most of the RP studies have considered either one or two nuclei for hyperfine interaction~\citep{ritz2000model,gauger2011sustained,bandyopadhyay2012quantum,timmel2001model,dellis2012quantum,pauls2013quantum,zhang2014sensitivity,carrillo2015environment}. RP studies with realistic number of nuclei in the hyperfine environment are limited~\citep{cai2010quantum,cintolesi2003anisotropic,lau2014alternative,solov2007magnetic,hiscock2016quantum}.
Moreover, the exact role of electronic coherence and possible interplay between coherence and compass parameters in a multi-nuclei RP system is still unclear. 

In this work, we investigate the role of coherence in avian magnetoreception by examining coherence dynamics for realistic compass parameters and relate it with the sensitivity of the compass. We relate the isotropy of the interactions and size of the nuclear spin space with the coherence of the compass. Additionally, we identify the compass parameter regime where its sensitivity is maximal and wherein the compass is therefore mostly likely to operate. We analyze the coherence dynamics for this parameter regime in order to understand the utility of coherence in the compass operation.

\section{Formalism of the Radical-Pair Model}
\label{RPModel}
The RP model captures the spin dynamics of the photo-generated radical pair interacting with the local geomagnetic field via the Zeeman interaction and with neighboring nuclei via the hyperfine interaction. 
It has been well studied for cases of either one or two nuclei interacting with electronic spins via hyperfine interaction~\citep{ritz2000model,gauger2011sustained,bandyopadhyay2012quantum,timmel2001model,dellis2012quantum,pauls2013quantum,zhang2014sensitivity,carrillo2015environment,poonia2015state,poonia2017functional}. However, the cryptochrome protein based RP system,  widely believed to be the actual protein responsible for radical pairs in the avian compass, can have multiple nuclear spins interacting with the radical pair via the hyperfine environment~\citep{ritz2000model,moller2004retinal}. Studies on the realistic RP system are few~\citep{cai2010quantum,cintolesi2003anisotropic,lau2014alternative,solov2007magnetic,hiscock2016quantum}. In this work, we investigate the multi-nuclei RP system. 
For an RP system $[A^{+} \enskip B^{-}]$, the Hamiltonian looks like:
\begin{eqnarray}
\begin{aligned}
    \hat{H} &= \hat{H}_A + \hat{H}_B \\
    \hat{H}_C &= \omega_0 (\vec{B}.\hat{S}) + \hat{H}_{C, hfi}\\
    \hat{H}_{C, hfi} &= \sum_{j = 1}^N a_{Cj}.\hat{S}_C \hat{I}_{Cj}
\end{aligned}
\label{RPHamiltonian}
\end{eqnarray}
where C $\in \{A,B\}$ and $\vec{B} = B cos\theta \hat{z} + B sin\theta cos \phi \hat{x} + B sin\theta sin\phi \hat{y}$ in Cartesian coordinates. $\omega_0$ denotes Larmor frequency, $a_{Cj}$ denotes hyperfine vector, and $\hat{I}_{Cj}$ denotes the nuclear spin of nucleus $j$ coupled to the electron in the radical C. For simplicity, we assume both dipole-dipole and exchange interactions are negligible.
The initial state of the radical pair is singlet state ($\ket{S}$) as both the radicals originate from a single molecule. After the formation of the radical pair, the ensuing evolution induces transitions between the singlet and triplet states. The compass operation is examined using the so-called singlet yield which is the fraction of the final chemical products coming from the singlet radical pairs after recombination. The Haberkorn model is employed to capture the entire dynamics~\citep{haberkorn1976density}. In this model, the singlet yield as a function of time is given as:
\begin{eqnarray}
\label{eq1}
\Phi_S(t) = \int_0^t k_S\rho_S(\tau) d\tau
\end{eqnarray}
where $\rho_S(t)$ is the fraction of singlet state at time t and $k_S$ denotes recombination rate through the singlet channel~\cite{tiersch2012decoherence,gauger2011sustained}\textcolor{red}. Fraction of singlet state at time t ($\rho_S(t)$) is given by the following:
\begin{eqnarray}
\label{Eq_RhoSt}
\begin{aligned}
\rho_S (t) &= Tr[\rho(t) \hat{Q}_S] \\
		&= \frac{1}{N} Tr[e^{\frac{-i}{\hbar}\hat{H}t} \hat{Q}_S e^{\frac{i}{\hbar}\hat{H}t} \hat{Q}_S]
\end{aligned}
\end {eqnarray}
where $\hat{Q_S}$ is the singlet projection operator, given as: $\hat{Q_S} = \frac{1}{4}\hat{I} - \hat{S_A},\hat{S_B}$ and $N = N_1 N_2$. $N_i$ is the size of nuclear spin space around radical $i$. Thus $N$ is the total size of the spin space of all nuclear spins interacting with electronic spins via hyperfine interaction. 
	The evolution of the radical pair density matrix may be obtained from a phenomenological master equation which models the recombination of the singlet and triplet radical pairs as anti-commutator terms. The equation goes as:
\begin{eqnarray}
\dot{\rho}(t) &= - \frac{i}{\hbar}[\hat{H}, \rho(t)] - \frac{k_S}{2}\{\hat{Q}_S, \rho(t)\} - \frac{k_T}{2}\{\hat{Q}_T, \rho(t) \}
\end{eqnarray}
where $k_S$ and $k_T$ are the recombination rates corresponding to the singlet and triplet radical pairs. For the popular choice of $k_S$ = $k_T$ = $k$, the equation can further simplifies to:
\begin{eqnarray}
\begin{aligned}
\dot{\rho}(t) & = - \frac{i}{\hbar} [\hat{H}, \rho(t)] - \frac{k}{2}\{\hat{Q}_S, \rho(t)\} - \frac{k}{2}\{{\hat{Q}_T}, \rho(t) \} \\ 
			 & = - \frac{i}{\hbar}[\hat{H}, \rho(t)] - k \rho(t)
\end{aligned}
\label{eq3}
\end{eqnarray}

We see that the time evolution of the density matrix here may be written simply as:
\begin{eqnarray}
\label{eq4}
\rho(t) = e^{-\frac{-i\hat{H}t}{\hbar}} \rho(0) e^{\frac{i\hat{H}t}{\hbar}} e^{-kt}
\end{eqnarray}
Thus, the density matrix time evolution is unitary Hamiltonian evolution multiplied by an exponential decay term.
Substituting the initial density matrix of the radical pair (singlet state) in the above equation, we have:
\begin{eqnarray*}
\begin{aligned}
\rho(t) &= \frac{1}{N} e^{-i\hat{H}t} \hat{Q}_S e^{i\hat{H}t} e^{-kt} \\
\rho (t) &= \frac{1}{N} e^{-i\hat{H}t}(\frac{1}{4} \hat{I} - \hat{S}_A . \hat{S}_B) e^{i\hat{H}t} e ^ {-kt}\\
 &= \frac{1}{N} e^{-i\hat{H}t}(\frac{1}{4}\hat{I} - \hat{S}_{Ax} \hat{S}_{Bx} - \hat{S}_{Ay} \hat{S}_{By} - \hat{S}_{Az} \hat{S}_{Bz}) e^{i\hat{H}t} e ^ {-kt}\\
 &= \frac{1}{4N} e^{-kt} - \frac{1}{N} \sum_{p = x,y,z} e^{-\frac{-i\hat{H}t}{\hbar}} \hat{S}_{Ap} \hat{S}_{Bp}e^{-\frac{i\hat{H}t}{\hbar}} e^{-kt}
\end{aligned}
\end{eqnarray*}
Now noting that the Hamiltonians of the two radicals ($\hat{H}_A$ and $\hat{H}_B$) commute in the absence of inter-radical spin interactions, the above equation transforms as:
\begin{eqnarray}
\begin{aligned}
\label{Eq_rhotime}
\rho (t) = \frac{1}{4N} e^{-kt} - \frac{1}{N} \sum_{p = x,y,z} e^{-\frac{-i\hat{H}_a t}{\hbar}} \hat{S}_{Ap}e^{\frac{i\hat{H}_a t}{\hbar}} \\ e^{-\frac{-i\hat{H}_b t}{\hbar}}\hat{S}_{Bp} e^{-\frac{i\hat{H}_bt}{\hbar}}  e^{-kt} \\
= \frac{1}{4N} e^{-kt} - \frac{1}{N} \sum_{p = x,y,z}  \hat{S}_{Ap}(t) \hat{S}_{Bp}(t)  e^{-kt}
\end{aligned}
\end{eqnarray}
We will use Eq.\eqref{Eq_rhotime} to calculate the dynamics of the electron spins in the next section. 
Moreover one should note that the effective evolution space has been reduced to the the individual radical Hilbert spaces, thereby reducing the computational complexity. This property will be useful in calculating the density matrix evolution involving more than one nuclear spin.\\
By substituting this $\rho(t)$ in Eq.~\ref{Eq_RhoSt}, the singlet population ($\rho_S (t)$) turns out to be:
\begin{eqnarray}
\label{Eq_RhoStFinal}
\rho_S (t) &= \frac{1}{4} + \frac{1}{N}\sum_{p = x,y,z} \sum_{q = x,y,z} R_{A pq} R _{B pq}
\end{eqnarray}
where,
\begin{eqnarray}
R_{A pq} = Tr[\hat{S}_{Ap} e^{\frac{-i}{\hbar}\hat{H}_{At}}\hat{S}_{Aq} e^{\frac{+i}{\hbar}\hat{H}_{At}}]
\end{eqnarray}
The singlet yield can now be calculated using Eq.~$\eqref{eq1}$. This gives us the following closed form expression for the singlet yield \cite{till1998influence}:
\begin{eqnarray}
\begin{aligned}
\Phi_S = \frac{1}{4} + \frac{1}{M}\sum_{p = x,y,z} \sum_{q = x,y,z} \sum_m \sum_n \sum_r \sum_s (\hat{S}_{Ap})_{mn} (\hat{S}_{Aq})_{nm} \\ 
(\hat{S}_{Bp})_{sr} (\hat{S}_{Bq})_{rs} \frac{k^2}{k^2 + ({\omega_{Am} - \omega_{An} + \omega_{Bs} - \omega_{Br}})^2}
\label{Eq_SY}
\end{aligned}
\end{eqnarray}
Henceforth, it is this expression that will be used to calculate the singlet yield. 

\begin{figure}[t]
\centering
\includegraphics[width=\linewidth]{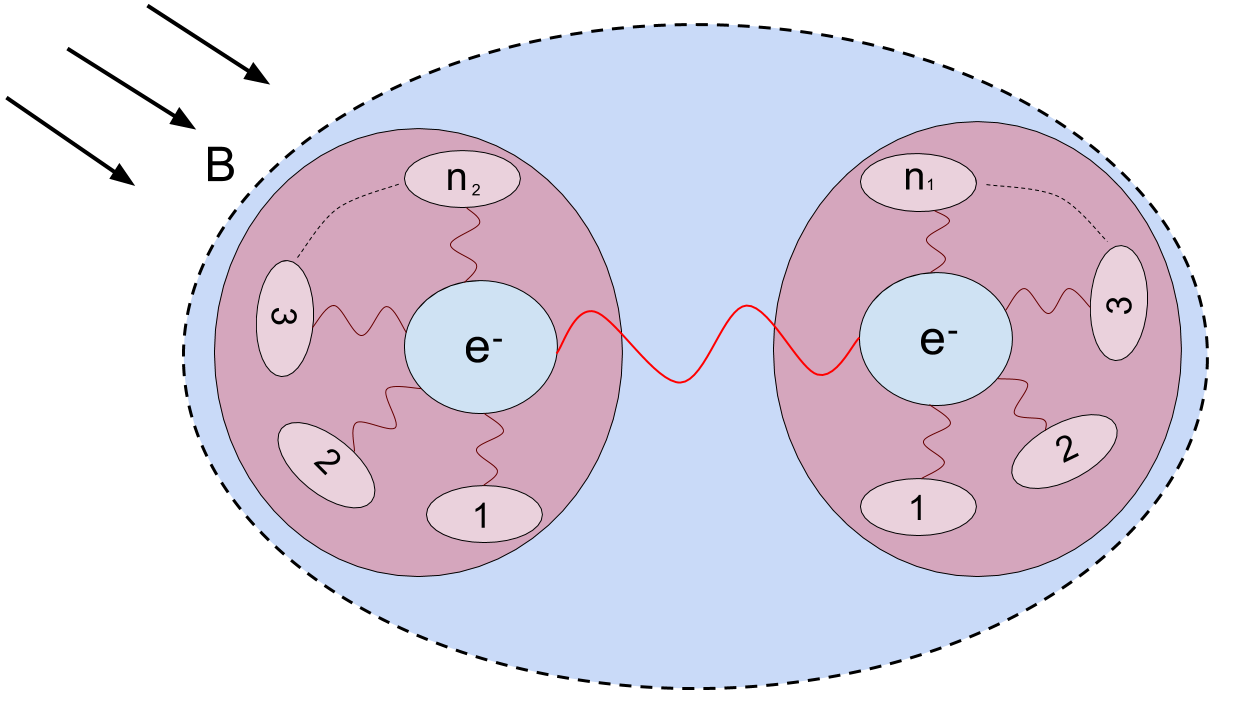}
\caption{ (Color online) A multi-nuclei radical pair system where electron on each radical is coupled to multiple nuclei.
Here, the red curl denotes correlation between the electron spins and the black curls denote hyperfine interactions between the electrons and nuclei.}
\label{MultiNucleiRPModel}
\end{figure}

\section{Results}
\label{SensCoh}
In order to elucidate the functioning of the compass, we investigate the dynamics of the singlet yield (i.e. the total population decay via the singlet channel) of the compass.
The singlet yield is calculated using Eq.~\ref{Eq_SY}, as derived in section~\ref{RPModel}.
To probe its directionality, we define the sensitivity of the compass as the difference between the maximum and minimum of the singlet yields with respect to the geomagnetic field inclination~\cite{bandyopadhyay2012quantum}:
\begin{eqnarray}
\label{Eq_Sensitivity}
\Delta_S = \Phi^{max}_S - \Phi^{min}_S
\end{eqnarray}
Thus, the sensitivity quantifies the directional dependence of the avian magnetoreception process and is a measure of the compass action itself.
We explore the sensitivities for various regimes of hyperfine interaction parameters and number of nuclei coupled to the individual electron spins in a radical pair.
Additionally, in order to quantify the coherence in the system, we use the 'relative entropy of coherence' as a measure of coherence~\cite{plenioquantifying}. 
It is defined as:
\\
\begin{eqnarray}
\label{Eq_CohQuantifier}
C(\rho) = S(\rho_{diag}) - S(\rho)
\end{eqnarray}
where $\rho$ is the density matrix of the system and $\rho_{diag}$ is the density matrix without the off-diagonal terms. $S(\rho)$ denotes von Neumann entropy of the system and is given by $-Tr(\rho ln(\rho))$. The Von Neumann entropy is zero for a pure state and has a maximum value of $ln(d)$ for a maximally mixed state where $d$ is the dimension of the spin Hilbert space.
The difference of the two von Neumann entropies is a measure of off-diagonal terms and physically corresponds to the coherence~\cite{plenioquantifying}.

In order to investigate the utilitarian role of coherence in the functioning of the avian compass, we study the coherence dynamics for a multi-nuclei radical pair system and correlate it with the sensitivity of the compass. We consider the cryptochrome based $\left[{FAD^{.-}} - Trp^{.+} \right]$ radical pair which is widely believed to be the system involved in avian magnetoreception~\citep{solov2007magnetic}. In our calculations, we have considered the $n-n$ radical pair system with $n=1,2,3$.
These systems avoid the shortcomings of the single-nucleus radical pair systems and capture the essential spin dynamics of the realistic multi-nuclei radical pair systems thus allowing us to make predictions about the cryptochrome based actual radical pair system that might have as much as 14 nuclei interacting with both the radicals~\cite{hiscock2016quantum}. The hyperfine parameter values taken for the calculations are given in Table~\ref{Table_HFValues}. 

\begin{figure}[t]
\centering
\begin{subfigure}[b]{\linewidth}
\includegraphics[width=\linewidth]{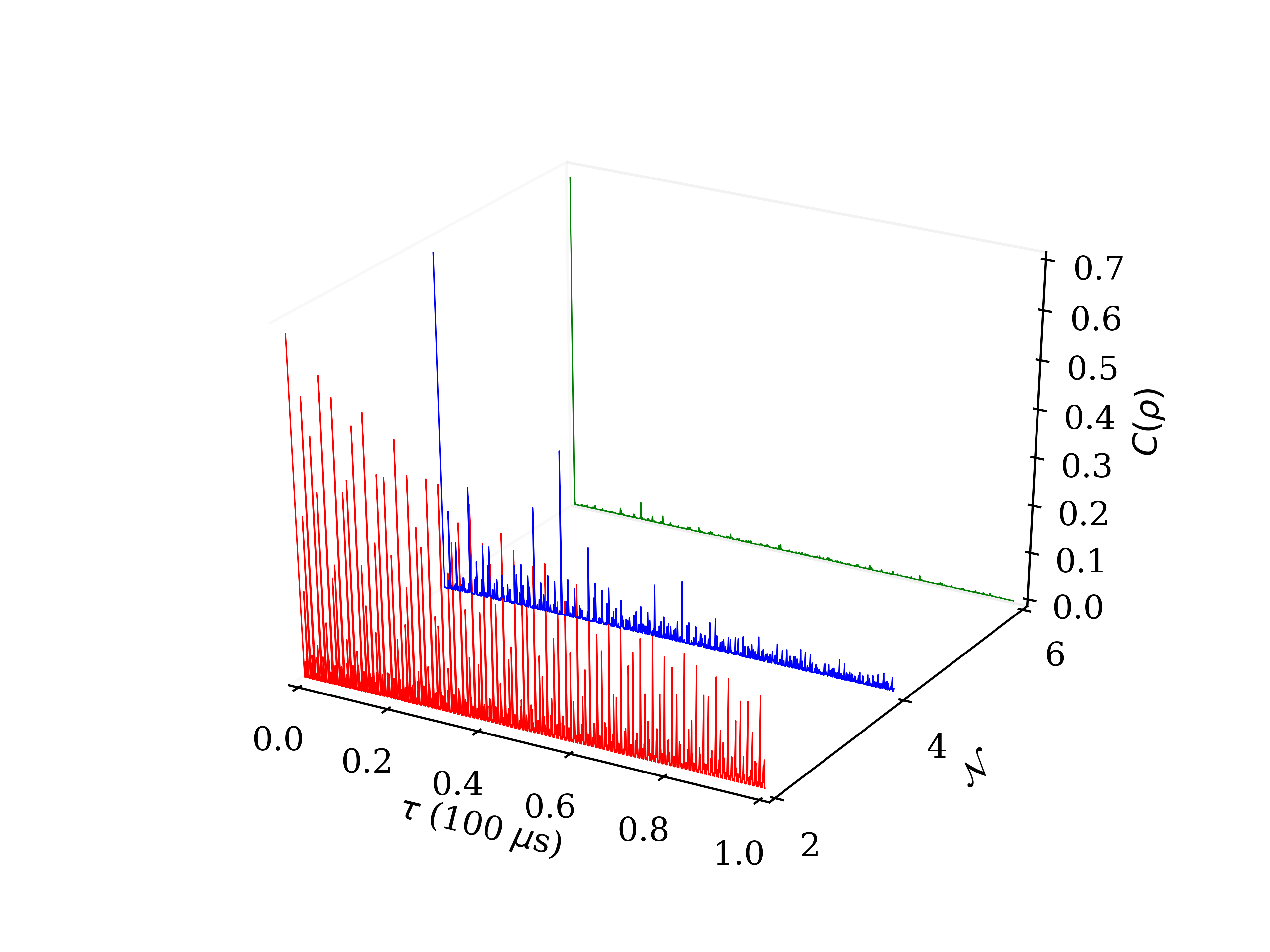}
\caption{}
\end{subfigure}
\begin{subfigure}[b]{\linewidth}
\includegraphics[width=\linewidth]{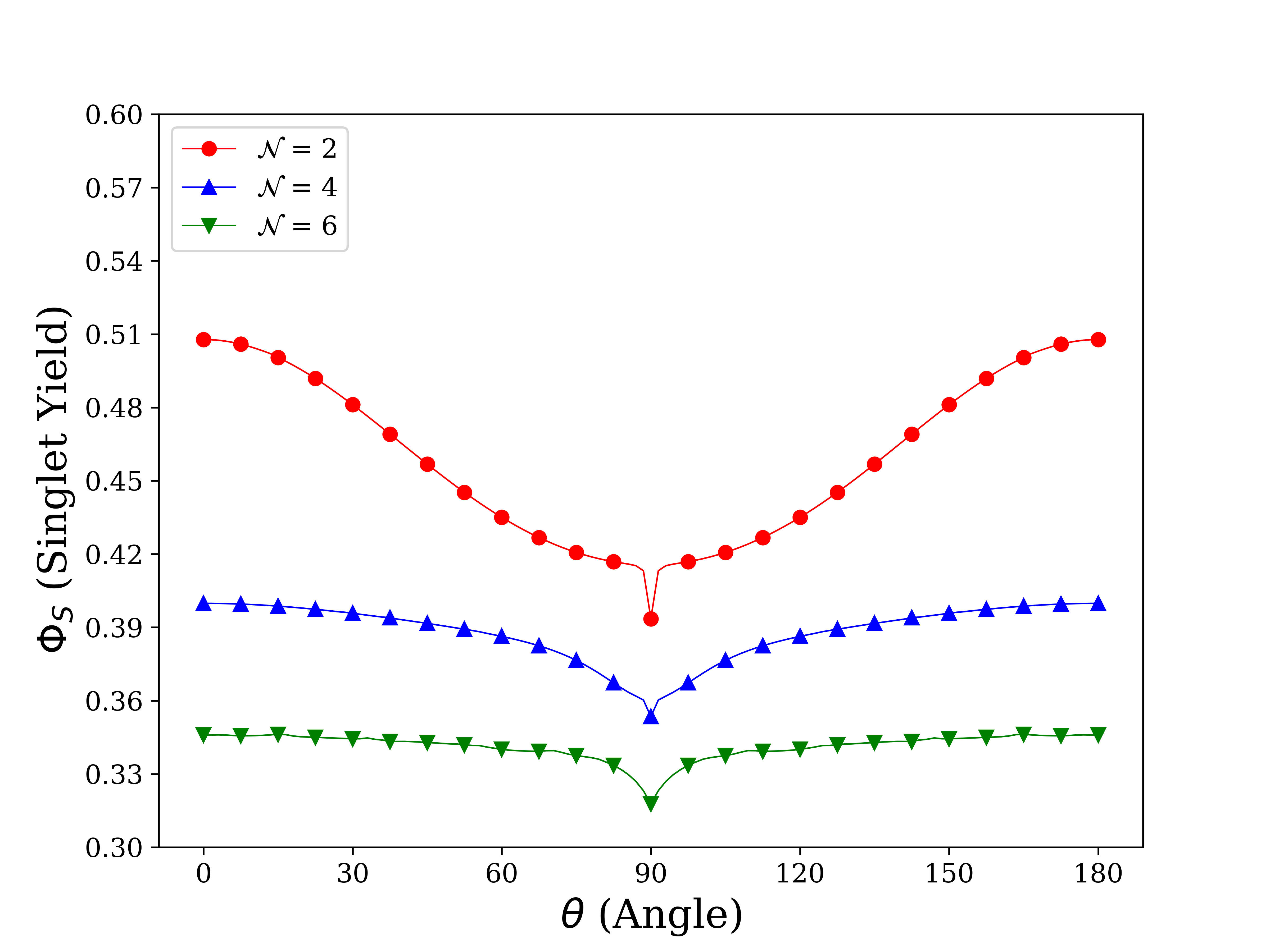}
\caption{}
\end{subfigure}
\caption{ (Color online) (a) Electronic coherence ($C(\rho)$) of the radical pair system as a function of time ($\tau$) for $\theta$ = 0 when multiple nuclei are coupled to the radical pair via hyperfine interaction. $C(\rho)$ is the relative entropy of coherence and is a quantifier of electronic coherence in the system. The plot describes the coherence dynamics of the system for 2, 4 and 6 nuclei interacting with the radical pair with recombination rates ($k_S$ = $k_T$) of $10^4$ $s^{-1}$. (b) The singlet yield as a function of geomagnetic field inclination for the aforementioned radical pair systems. 
For these plots, the hyperfine coupling strengths are given in Table
~\ref{Table_HFValues}.\\
}
\label{Fig_CohSens_NoOfNuclei}
\end{figure}

\begin{table}[t]
\centering
\begin{tabular}{ |p{2cm}||p{1.75cm}|p{1.75cm}|p{1.75cm}|  }
\hline
\multicolumn{4}{|c|}{Hyperfine couplings for $FAD^{.-}$} \\
\hline
Nuclei & $a_x(mT)$ &$a_y(mT)$&$a_z(mT)$\\
\hline
N5   &-.0989& -.0989& 1.7569\\
N10 & -.0241& -.0144& .6046  \\
H6 &-.5304&-.4336&-.1976\\
\hline
\multicolumn{4}{|c|}{Hyperfine couplings for $Trp^{.+}$} \\
\hline
Nuclei & $a_x(mT)$ &$a_y(mT)$&$a_z(mT)$\\
\hline
N1  &0&0&1.0812\\
H1&0.4716&-.36990&0  \\
H4& -.74&-.536&-.1879 \\
\hline
\end{tabular}
\caption{Hyperfine interaction parameter values for $\left[{FAD^{.-}} - Trp^{.+} \right]$ radical pair system. Values are taken from ref.~\cite{hiscock2016quantum}.}
\label{Table_HFValues}
\end{table}

We observe from our calculations that the coherence decays due to population decay and interactions with the coupled nuclear spins. 
The effect of these factors is captured in Eq.\eqref{Eq_rhotimelb} where the exponential factor describes the population decay and the other term models the evolution due to hyperfine and Zeeman interactions.
\begin{eqnarray}
\label{Eq_rhotimelb}
\rho(t) = \underbrace{(\frac{1}{4N}  - \frac{1}{N} \sum_{p = x,y,z}  S_{Ap}(t) S_{Bp}(t)  )}_\text{Evolution}  \underbrace{e^{-kt}}_\text{Population decay}
\end{eqnarray}
The coherence decay due to nuclear interactions further depends on two factors: i) number of the nuclei interacting with the radical pair, ii) isotropy of the hyperfine interaction tensor. 

We analyze their effects separately in the following manner. 
The initial state (at t = 0) of the radical pair is the coherent singlet state. First, in order to investigate the effect of number of nuclei on electronic coherence,
we plot the coherence (as quantified by Eq.~\ref{Eq_CohQuantifier}) dynamics of the radical pair system with 2, 4, and 6 nuclei interacting with the radicals (1,2 and 3 nuclei interacting with each radical) 
for the hyperfine tensor of $(a_x, a_y, a_z)$
from the Table ~\ref{Table_HFValues}. The results are shown in Fig~\ref{Fig_CohSens_NoOfNuclei} where part (a) of the figure shows time evolution of coherence as the number of nuclei is increased and part (b) shows the singlet yield as a function of geomagnetic field inclination for the corresponding hyperfine parameters and number of nuclei. The oscillations in the coherence have been explained in Appendix A. The hyperfine parameters in this case are chosen as follows -- two nuclear spin system includes contribution from the first nucleus of $FAD^{.-}$ and $Trp^{.+}$ each as mentioned in the Table~\ref{Table_HFValues}. Similarly four and six nuclear spin system includes contributions from the first two and first three nuclear spins of $FAD^{.-}$ and $TrpH^{.+}$ radicals as mentioned in the Table~\ref{Table_HFValues}. The figure clearly demonstrates that as the number of nuclei are increased from 2 to 6, the coherence vanishes very fast and singlet yield flattens, thereby decreasing the compass sensitivity. This is expected because more nuclei amounts to a bigger bath interacting with the radical pair spin, hence causing it to decohere fast.
This makes it unlikely for coherence to be sustained in the case of recent calculations done with a total of 14 nuclear spins coupled to the radical pair~\cite{hiscock2016quantum} as these calculations included the same couplings we took into account for our calculations.

\begin{figure}[b]
\centering
\begin{subfigure}[b]{\linewidth}
\includegraphics[width=\linewidth]{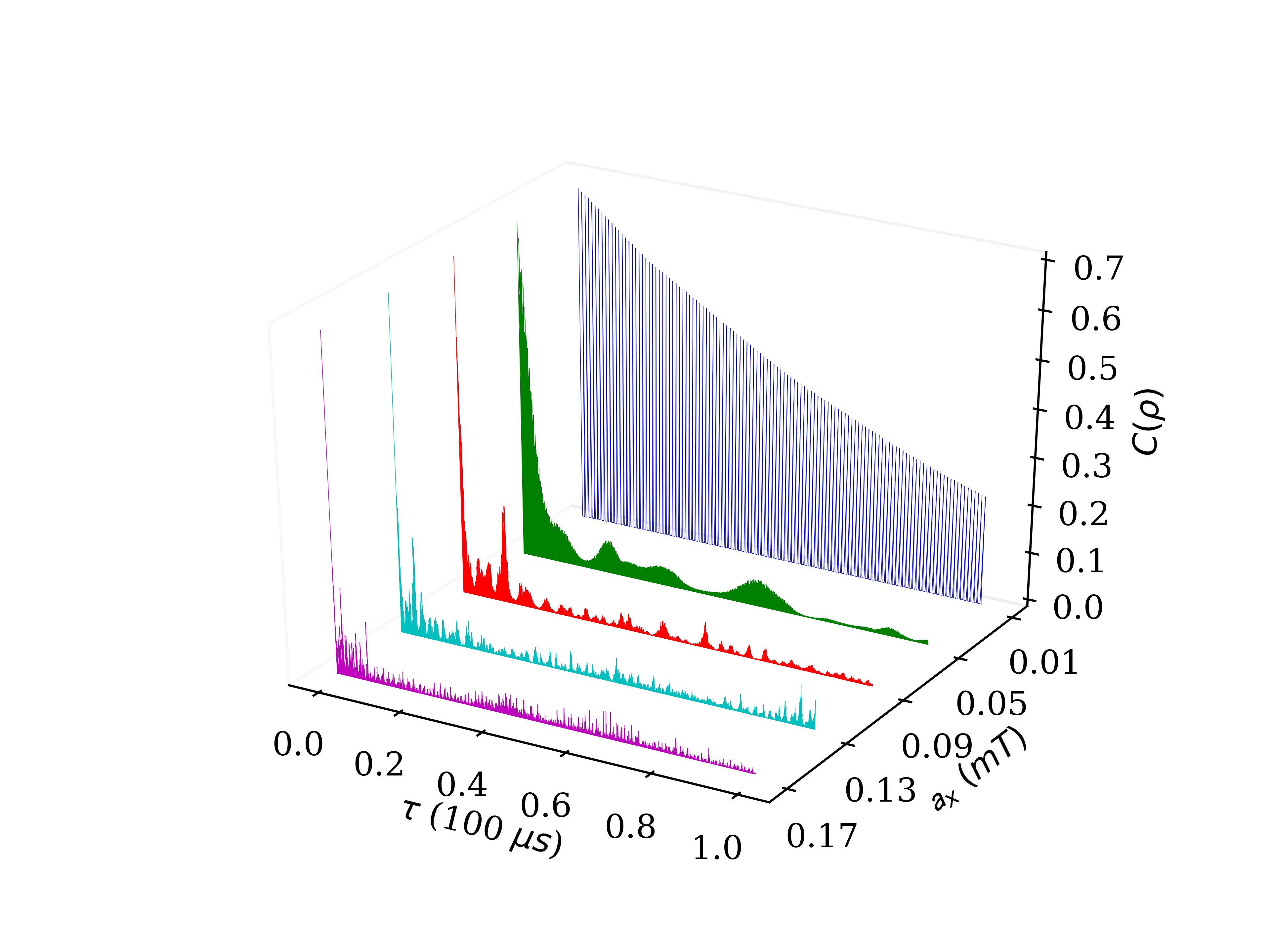}
\caption{}
\end{subfigure}
\begin{subfigure}[t]{\linewidth}
\includegraphics[width=\linewidth]{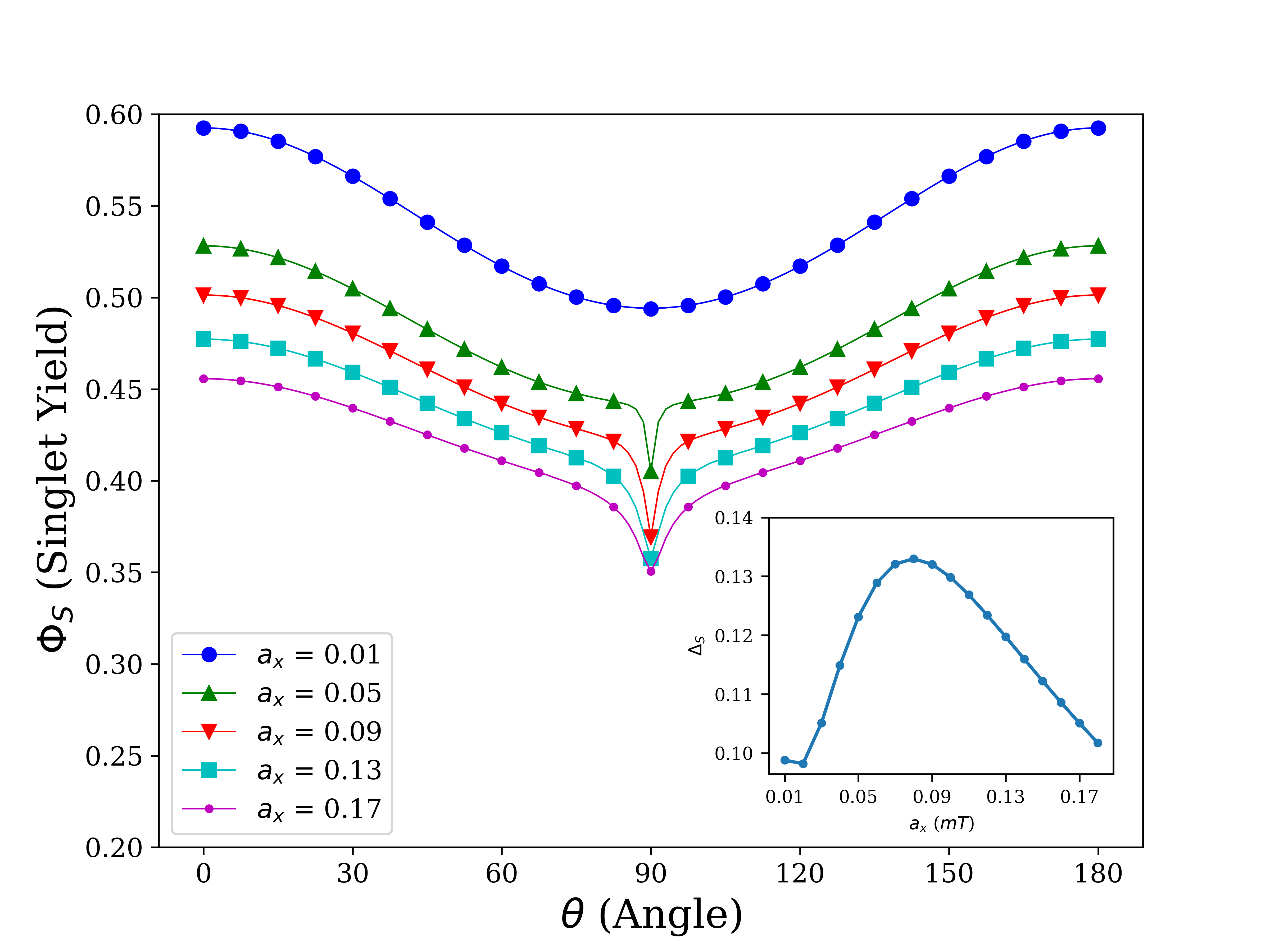}
\caption{}
\end{subfigure}
\caption{ (Color online) (a) The radical pair coherence dynamics ($\theta$ = 0) for various hyperfine interaction strengths. This plot describes the effect of isotropic hyperfine couplings on the coherence in the system. Here we vary $a_x$ = $a_y$ components of the hyperfine tensor with $a_z$ = 1.0812 mT
for $N_1$ of $Trp^{.+}$ and recombination rates ($k_S$ = $k_T$) as $10^4$ $s^{-1}$.
(b) Effect of hyperfine tensor anisotropy on the singlet yields for the hyperfine interaction parameters used in part (a) of the figure. The inset shows the sensitivity (difference between max. and min. singlet yield) as a function of $a_x (a_y)$.
}
\label{Fig_CohSens_HFIsotropy}
\end{figure}

Second, in order to observe the effect of isotropy of hyperfine tensor on electronic coherence of the radical pair system, we calculate the coherence dynamics for $\theta = 0$ (See Appendix~\ref{appendixb} for variation of coherence dynamics with $\theta$) and singlet yields for a 3-3 radical pair system for different values of the transverse hyperfine constants. 
In these calculations, we consider a case where the electron on each radical is coupled with three nuclear spins. In order to simplify the interpretation, we calculate the dynamics of the radical pair assuming all three hyperfine interaction strengths to be equal. 
We take the axial hyperfine constant ($a_z = $1.0812 mT )  of N1 of $Trp^{.+}$ and vary the transverse components ($a_x$ and $a_y$) from 0 to 0.17 mT.
The results are presented in Fig.~\ref{Fig_CohSens_HFIsotropy}. As is clear from Fig.~\ref{Fig_CohSens_HFIsotropy} (a), the coherence is sustained for a completely anisotropic hyperfine constant ($a_x = a_y = 0$) and starts deteriorating as we increase the isotropy of the hyperfine tensor (by increasing the values of $a_x$ and $a_y$). The coherence decays precipitously for $a_x = a_y = 0.17$ mT and beyond.

Now, we attempt to further clarify the effect of the hyperfine isotropy on coherence. Our starting point is the radical pair Hamiltonian (Eq.\eqref{RPHamiltonian}):
\begin{align*}
\hat{H} &= \omega_0 (\vec{B}.\hat{S}) + \sum_{C} \sum_{j = 1}^N a_{Cj}\hat{S}_C . \hat{I}_{Cj}
\end{align*}
For these calculations, we assume $\phi = 0$. For $\theta$ = 0 and completely anisotropic hyperfine coupling($a_{Cjx}, a_{Cjy} = 0$), the total Hamiltonian can be written as:
\begin{align*}
\hat{H} &=  \hat{S_z}(\omega_0 B_z + \sum_{C} \sum_{j = 1}^N a_{Cjz} \hat{I}_{Cjz} )\\
\end{align*}
Note that in this case the nuclear spin space can be uncoupled from electron spin space for $\theta$ = 0. Hence, this case of completely anisotropic hyperfine coupling, viz. ($a_{Cjx}, a_{Cjy} = 0$), only leads to an effective field and no decoherence due to the nuclear environment, thereby sustaining coherence for $\theta$ = 0 irrespective of the number of nuclei coupled to the electron pair. 
This result can be approximated to all the angles provided $\gamma B$ is much smaller than the anisotropic hyperfine coupling with the nuclear spins ($a_z$ in this case) which is  the case when we consider realistic couplings, as given in the Table ~\ref{Table_HFValues}.
\begin{figure}[t]
\centering
\begin{subfigure}{\linewidth}
\includegraphics[width=\linewidth]{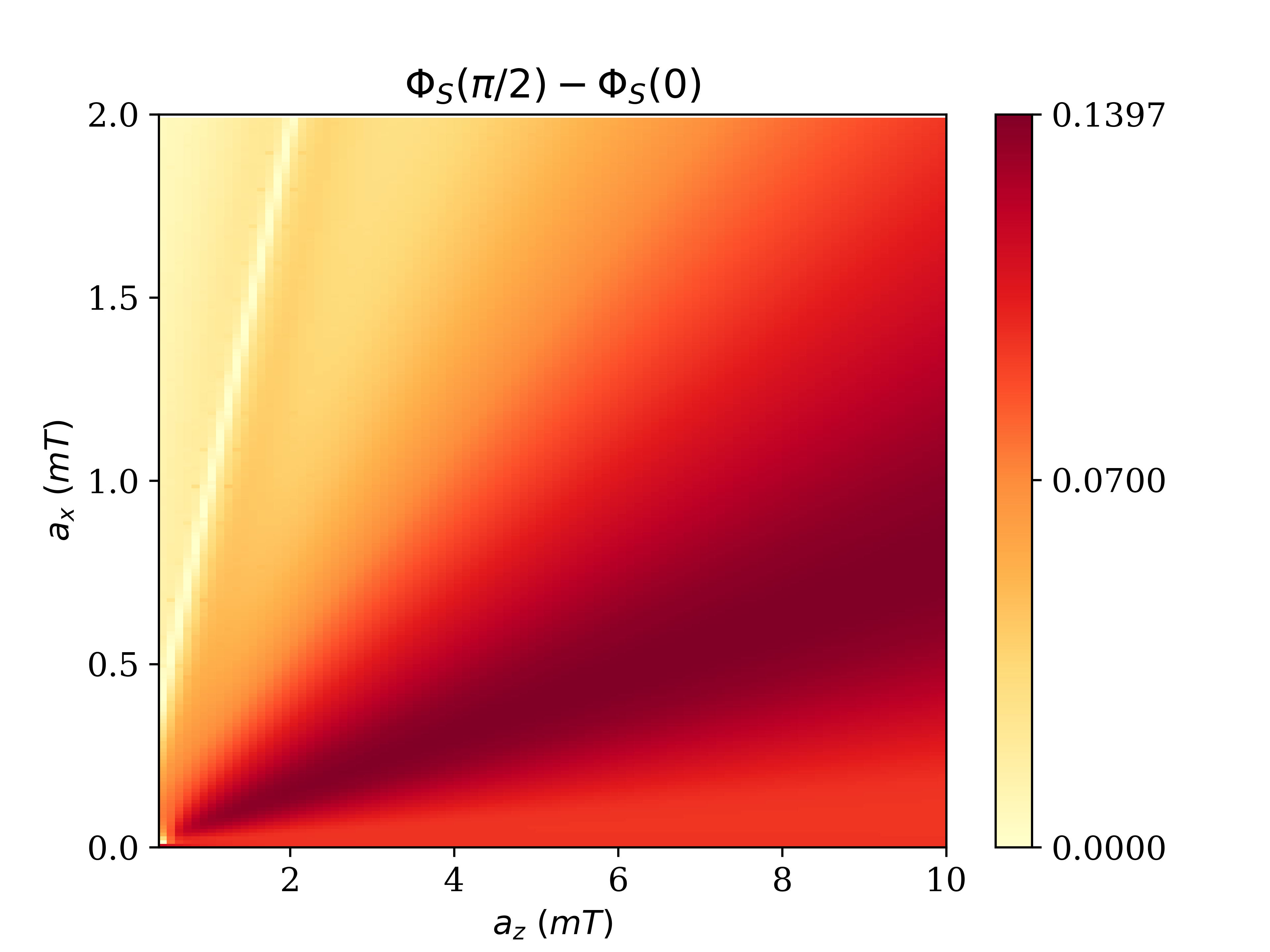}
\caption{
}
\end{subfigure}
\begin{subfigure}{\linewidth}
\includegraphics[width =\linewidth]{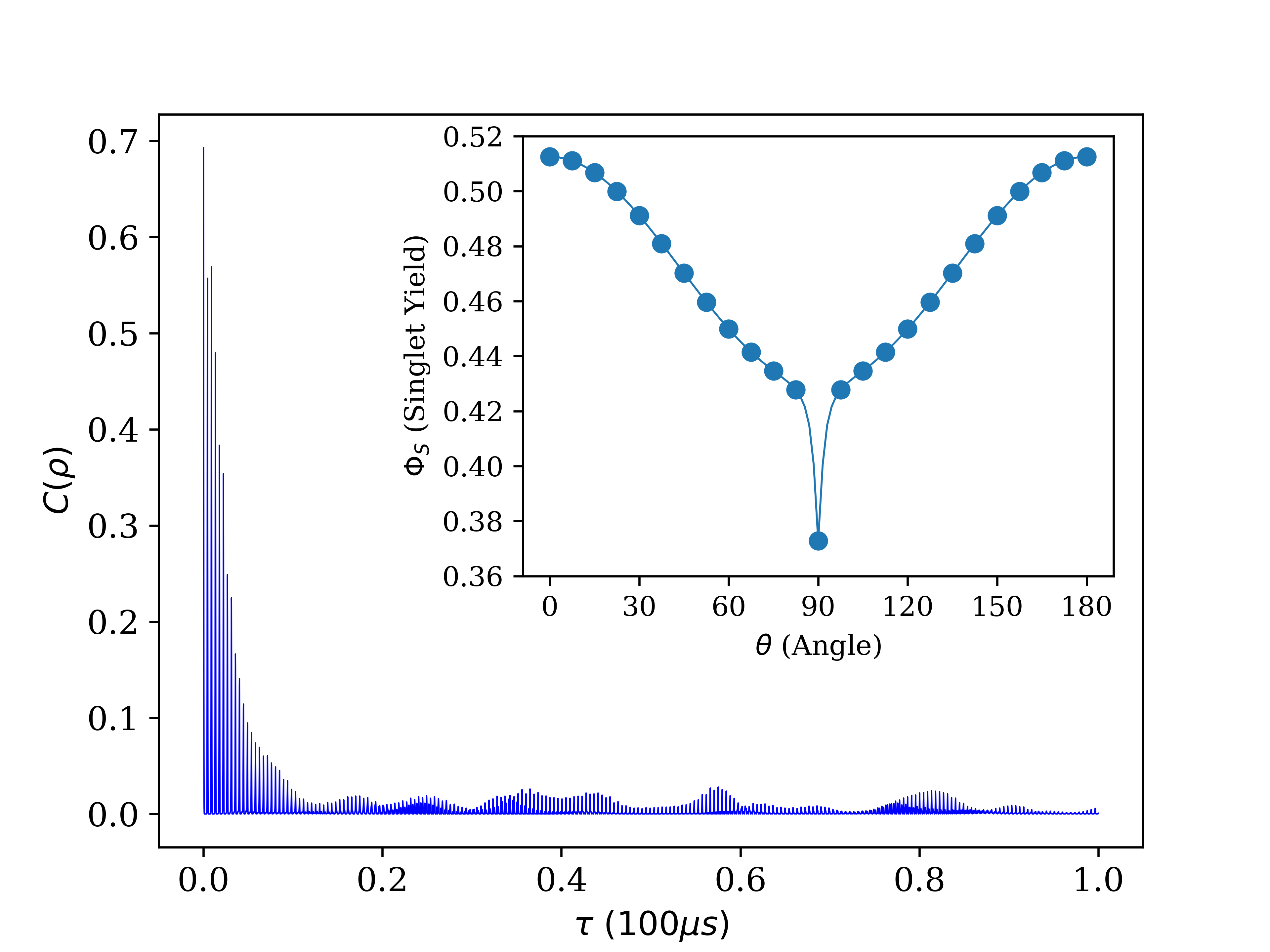}
\caption{}
\end{subfigure}
\caption{(Color online)  Difference between singlet yields at $\theta$ = $0^o$ and $90^o$  and coherence dynamics for different regimes of hyperfine parameter for a six nuclei radical pair system. a) Compass sensitivity as a function of longitudinal and transverse components of the hyperfine parameter. Here the recombination rates ($k_S$ = $k_T$) are $10^4$ $s^{-1}$. b) Coherence dynamics for hyperfine parameters of $(a_x,a_y,a_z) = (0.16,0.16,2.0) mT$. This parameter lies on the maximum sensitivity regime in figure (a). The inset shows the singlet yield profile for the needle parameters corresponding to the maximum difference in the singlet yield values.}
\label{Fig_Sens_HF}
\end{figure}

Interestingly, from Fig.~\ref{Fig_CohSens_HFIsotropy}, we notice that the sensitivity (shown in Fig.~\ref{Fig_CohSens_HFIsotropy} (b) inset) does not decay with an increase in the isotropy of the hyperfine tensor. On the other hand, the singlet yield develops a sharp dip at $\theta = 90 \deg$ (termed as "compass needle" in an earlier work~\citep{hiscock2016quantum}) thus increasing the sensitivity of the compass. This enhancement of compass sensitivity as the transverse hyperfine components are increased and attain a maximum (See Fig.~\ref{Fig_CohSens_HFIsotropy} inset), leads to the sharpest needle formation ~\cite{hiscock2016quantum}. However, this needle vanishes as we increase the transverse hyperfine components further, thus leading to a decline in sensitivities.
This concavity of the sensitivity in contrast to the decaying coherence
provides a hint on how compass sensitivity might endure despite low coherence in the RP system;
in other words, the avian compass need not have sustained coherence in order to be sensitive to the geomagnetic field. This seems to stand in contrast to a few earlier claims~\citep{gauger2011sustained,hiscock2016quantum}.
\begin{figure}[t]
\centering
\begin{subfigure}{\linewidth}
\includegraphics[width=\linewidth]{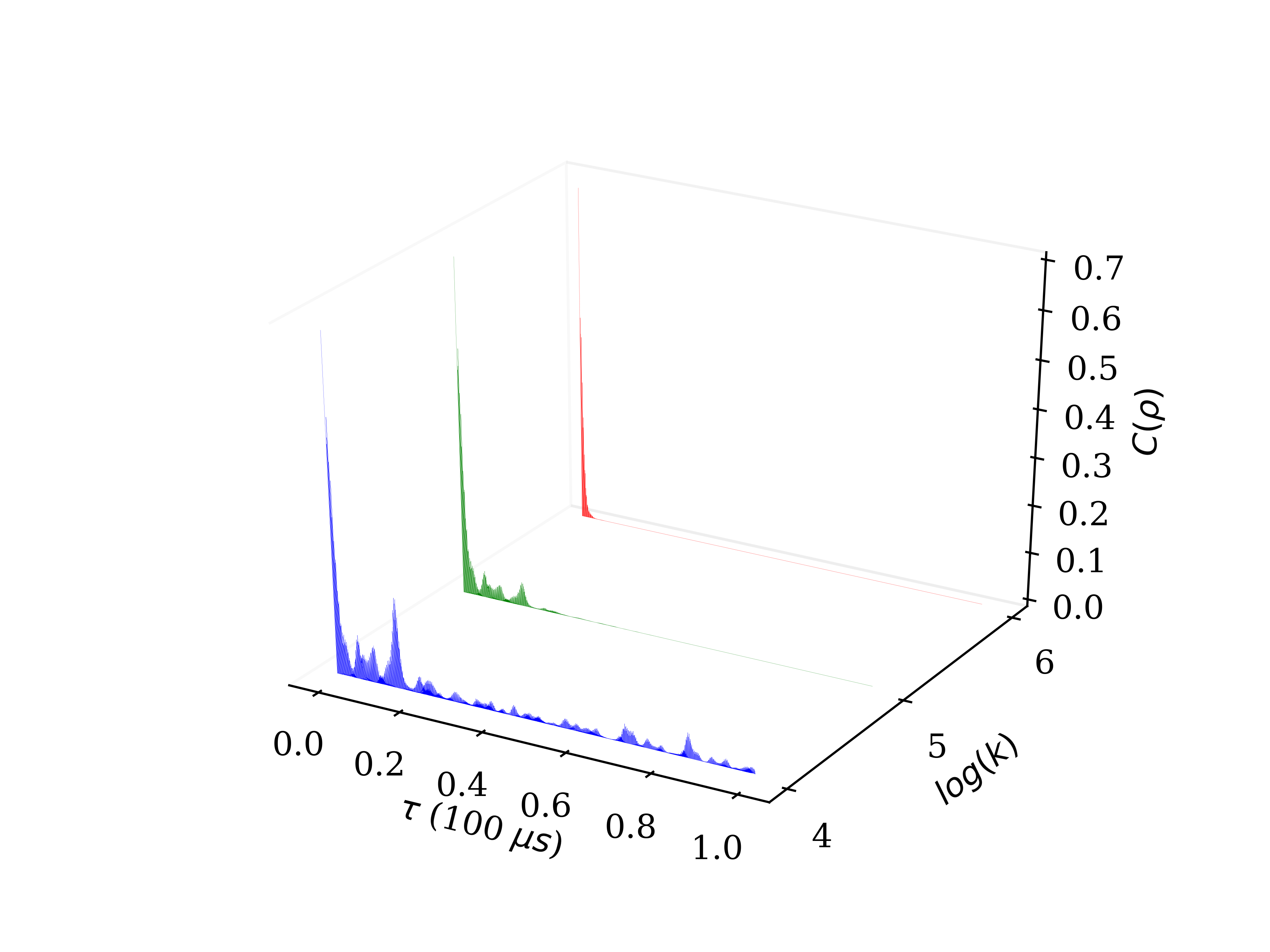}
\caption{}
\end{subfigure}
\begin{subfigure}{\linewidth}
\includegraphics[width=\linewidth]{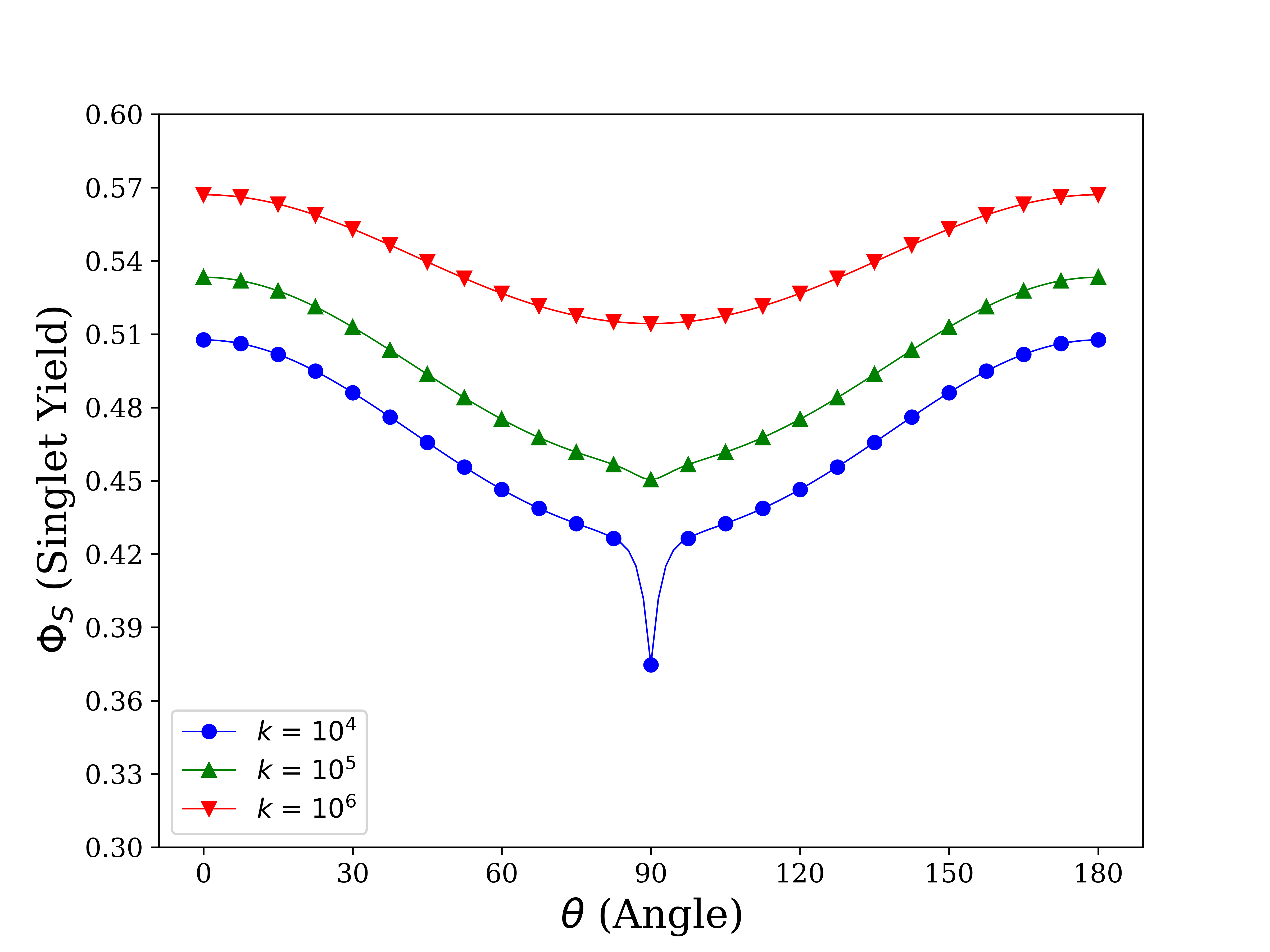}
\caption{}
\end{subfigure}
\caption{(Color online) Coherence dynamics (a) and the singlet yield profiles (b) for recombination rates ($k = k_S = k_T$) of of $10^4$, $10^5$, and $10^6 s^-1$ in the 3-3 radical pair system (3 nuclei interacting with each electron in the radical pair) with hyperfine parameters of (0.08,0.08, 1.0812) $mT$.}
\label{Fig_CohSens_RecombRates}
\end{figure}

Therefore we now move on to explore the compass parameter space in terms of both the longitudinal and axial hyperfine parameters; specifically the regime where it is most sensitive to the geomagnetic field and examine the coherence dynamics there. To that end, we plot the difference between singlet yields at $\theta = 0^{o}$ and $\theta = 90^{o}$ as a function of both axial and transverse hyperfine couplings for a radical pair system similar to our previous calculation with three nuclei coupled to each electron spin, cf. Fig.~\ref{Fig_CohSens_HFIsotropy}. The results are shown in Fig.~\ref{Fig_Sens_HF}. Here we observe that the compass does not have high sensitivity for a completely anisotropic hyperfine tensor where the coherence is maximal. Neither does it have the best sensitivity for an isotropic hyperfine tensor where the coherence does not persist at all. The sensitivity maxima is achieved for a disk-shaped hyperfine tensor ($a_z \gg a_x = a_x \neq 0$) which directly corroborates our earlier work~\citep{poonia2017functional}; note that this happens to be a regime where the coherence is not sustained. This may be seen from the coherence dynamics of electron spins, plotted in Fig.~\ref{Fig_Sens_HF} (b) where one can clearly observe that the coherence is not sustained for the parameter regime where the sensitivity is highest due to the needle in the singlet yield. This hints at the possibility that the avian compass may not be a system with quantum coherence after all.

Lastly, we calculate the effect of varying the radical pair recombination rates ($k_S$ and $k_T$) on coherence dynamics and sensitivity  of the avian compass. The results are shown in Fig.~\ref{Fig_CohSens_RecombRates} where we plot the coherence dynamics and singlet yields for different rate constants. The compass coherence dynamics is plotted for the recombination rates of $k_S = k_T = 10^4, 10^5$ and $10^6$.
As is evident from the figure, the coherence is suppressed as the recombination rates are increased. This is also accompanied by a drop in the compass sensitivity, as shown in Fig.~\ref{Fig_CohSens_RecombRates} (b). The needle in the singlet yield disappears for higher values of the recombination rates, and thereby
causing the sensitivity to drop.
It may be noted that in the context of this figure, all the nuclear spins have been considered identical for the calculation of the singlet yields and the coherence dynamics. 

\section{Discussion and Conclusion}
\label{Conclusion}
In summary, we have presented a formalism for calculating the coherence dynamics and singlet yields for a multi-nuclear radical pair model. We would like to point out here that our analysis is limited to a radical pair model comprising 6 nuclei, while the real system based on the cryptochrome protein comprises even more. While we are able to capture the essential coherence dynamics of a realistic radical pair system, our conclusions would bear verification by calculations on the full cryptochrome based system. Our analysis of the coherence dynamics for a range of hyperfine and recombination rate parameters lets us identify the parameter regime where the avian compass would be maximally sensitive to the local geomagnetic field. Here the singlet yield as a function of the field inclination takes on a peculiar needle-like characteristic. Our calculations show that, for this optimal set of parameters, the compass can have high sensitivity despite low coherence. This leaves us with a basic question about the nature of the avian compass: is it truly governed by quantum coherence? Or have its parameters evolved over millenia just to make the radical pair spin dynamics most sensitive to field inclination? We suggest that the actual compass parameters need to be deduced from more in-vivo or in-vitro experiments on migratory avian species; comparison with the parameters presented here could help to fully clarify the role of coherence and the physical origin of the needle-like sensitivity characteristic.



\section{Acknowledgements}
\label{Acknowledgements}
We would like to thank Peter Hore for insightful communications. 
We are also grateful for the support from the Ministry of Electronics and Information Technology through the Centre of Excellence in Nanoelectronics at IIT Bombay.

\bibliography{main.bib}

\appendix

\section{Oscillations in coherence dynamics: State transitions point of view}
\label{appendixa}
The role of Zeeman and hyperfine interactions in inducing various spin transitions in order to make the radical pair spin dynamics magnetosensitive has been demonstrated in \cite{xu2014effect,poonia2015state}. These results are key to understanding the interplay of singlet and triplet spin states in making the radical pair spin dynamics magnetosensitive. Here, we explain the oscillations in coherence dynamics (Fig.~\ref{Fig_CohSens_NoOfNuclei} and ~\ref{Fig_CohSens_HFIsotropy} in the text) using these state transitions. In a one-nucleus radical pair system, the z component of hyperfine tensor is responsible for inducing the transitions between $S$ and $T_0$ states whereas x and y components induce transitions between $S$ and $T_{\pm}$. Zeeman terms are responsible for inducing transitions among ($T_0, T_+$ and $T_-$ states). A detailed account of the effect of various Hamiltonian terms (hyperfine and Zeeman) on the state transitions has been presented in ref.~\cite{poonia2015state}. This kind of spin-specific transitions induced by various Zeeman and hyperfine terms give rise to the oscillatory behavior that we observe in the coherence dynamics of the radical pair system, cf. Fig. ~\ref{Fig_CohSens_NoOfNuclei} and ~\ref{Fig_CohSens_HFIsotropy} in the text. The envelop of these plots is the measure of the coherence in the RP system.
In these figures, the hyperfine parameters are taken from Table~\ref{Table_HFValues} (given in the text). 
Depending on the value of these parameters in different directions, they induce transitions among various spin states with different frequencies. Additionally, depending on the inclination of the geomagnetic field ($\theta$), Zeeman field's x-component, z-component or a combination thereof might be contributing in the system Hamiltonian. z-component of Zeeman field does not induce any transition among the spin states whereas x-component induces $T_0 \leftrightarrow T_+$ and $T_0 \leftrightarrow T_-$ transitions. Therefore, at $\theta = 0^{\deg}$ (when x-component of Zeeman field is zero), only $S \leftrightarrow T_0$ transition is active. The relative entropy of coherence i.e. the coherence measure used here (described in more detail in the manuscript) is essentially a measure of the off-diagonal elements of the density matrix. In a nutshell, the inter-state spin transitions along with recombination of radicals create the oscillations as seen in Fig. ~\ref{Fig_CohSens_NoOfNuclei} and ~\ref{Fig_CohSens_HFIsotropy} in the relative entropy of coherence. For non-zero values of the geomagnetic field inclination, both x-component and z-component of the Zeeman field is non-zero. Hence, transitions are induced between all four spin states. This makes the density matrix change its diagonal elements more frequently. Therefore, the oscillatory dynamics is observed in the coherence measure.
\begin{figure}[t]
\centering
\includegraphics[width = \linewidth]{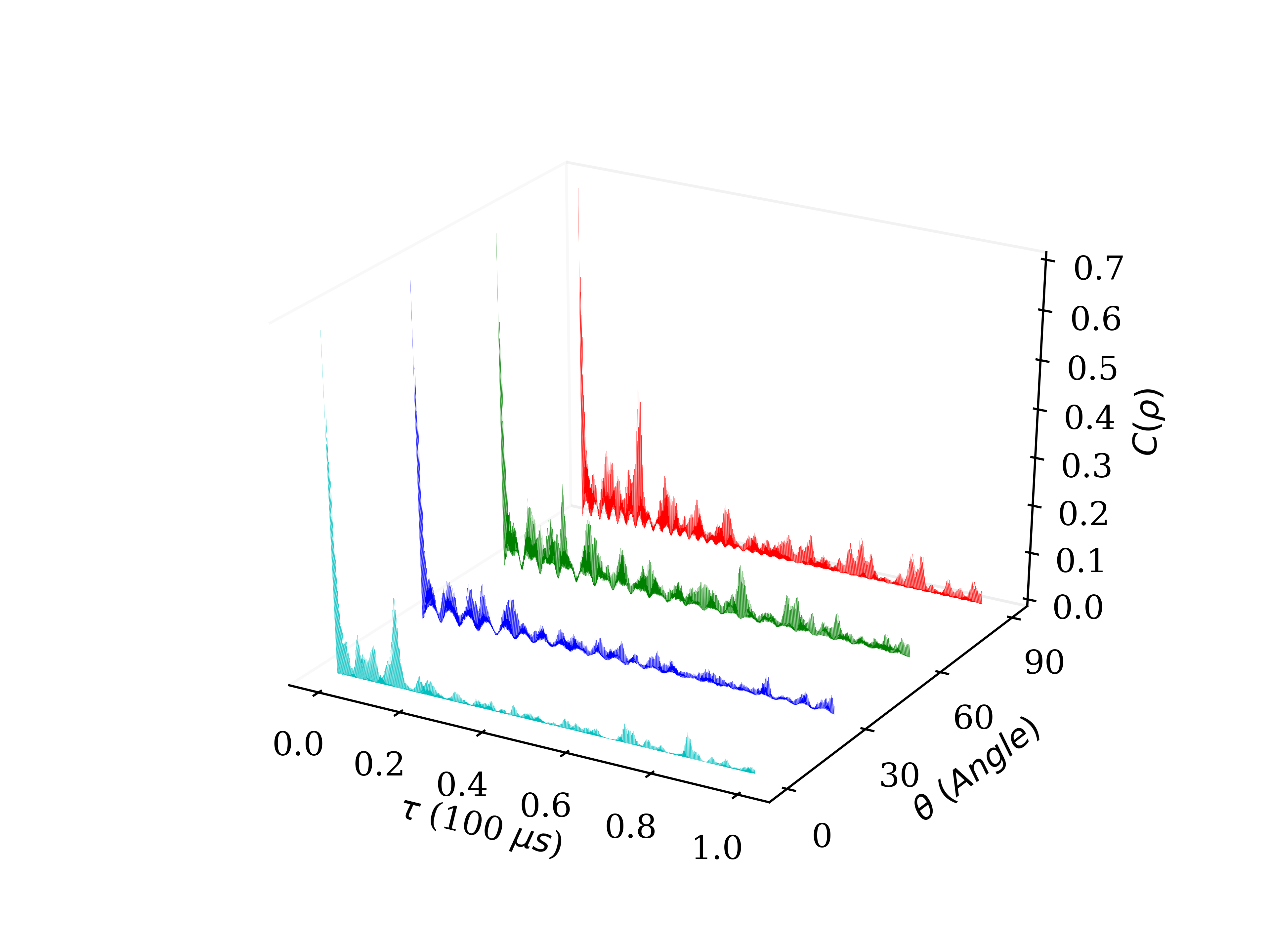}
\caption{(Color online) Coherence dynamics for various angles of the radical pair axis with the geomagnetic field inclination.
Here we consider the same parameters as in Fig.~\ref{Fig_CohSens_RecombRates}}
\label{FigSup1}
\end{figure}

\section{Coherence dynamics for various angles with the Zeeman field}
\label{appendixb}
The transitions of the electron spins depends on the nuclear environment and the angle with Zeeman field and hence cause oscillations in coherence dynamics (See Appendix ~\ref{appendixa}).
But the coherence decay (for example in Fig.~\ref{Fig_Sens_HF}) depends mainly on the terms described in Eq.~\eqref{Eq_rhotimelb}.
This is  illustrated in Fig.~\ref{FigSup1} where we calculate coherence dynamics for various angles of the axis of radical pair with the geomagnetic field inclination.

\end{document}